\documentclass[aps,pre,preprint,groupedaddress,showpacs]{revtex4-1}
\usepackage[dvips]{graphicx}
\usepackage{textcomp}
\usepackage{amssymb}

\newcommand{\mc}{\multicolumn}

\begin{document}

\title{
\Large\bf  The dynamic critical exponent $z$ of the three-dimensional Ising 
universality class: Monte Carlo simulations of the improved Blume-Capel model
  }

\author{Martin Hasenbusch}
\email[]{M.Hasenbusch@thphys.uni-heidelberg.de}
\affiliation{
Institut f\"ur Theoretische Physik, Universit\"at Heidelberg,
Philosophenweg 19, 69120 Heidelberg, Germany}
\date{\today}

\begin{abstract}
We study purely dissipative relaxational  dynamics in the three-dimensional Ising universality class. 
To this end, we simulate the improved Blume-Capel model on the simple cubic lattice
by using local algorithms.  We perform a finite size scaling analysis 
of the integrated autocorrelation time of the magnetic 
susceptibility in equilibrium  at the critical point. We obtain $z=2.0245(15)$
for the dynamic critical exponent. 
As a complement, fully magnetized configurations are suddenly quenched to the 
critical
temperature, giving consistent results for the dynamic critical exponent.
Furthermore, our estimate of $z$ is fully consistent with recent field theoretic results.
\end{abstract}

\maketitle

\section{Introduction}
In the neighborhood of a second order phase transition, thermodynamic 
quantities diverge, following power laws. For example, the correlation 
length $\xi$ diverges as 
\begin{equation}
\label{xipower}
\xi = f_{\pm} |t|^{-\nu}  \; \left(1 +a_{\pm}  |t|^{\theta} + b t + ...\right) \;,
\end{equation}
where $t=(T-T_c)/T_c$ is the reduced temperature and $\nu$ the critical 
exponent of the correlation length. 
The subscript $\pm$ of the
amplitudes $f_{\pm}$ and $a_{\pm}$ indicates the high ($+$) and the low
($-$) temperature phase, respectively.
Second order phase transitions are grouped into universality classes. For all
transitions within such a class, critical exponents like $\nu$ assume the identical
value.
These power laws are affected by corrections. There are non-analytic or confluent
and analytic ones. 
Also correction exponents such as $\theta=\omega \nu$ are universal.
Amplitudes such as
$f_{\pm}$, $a_{\pm}$ and $b$ depend on the microscopic details of the system.
However certain combinations, so called amplitude ratios, assume universal
values.
Universality classes are characterized by the symmetry properties of the
order parameter at criticality, the range of the interaction and the spacial
dimension of the system.
Currently the most accurate estimates of static critical exponents for the 
universality class of the three-dimensional Ising model are 
$\nu=0.6299709(40)$, $\eta=0.0362978(20)$, and the 
exponent of the leading correction $\omega=0.82968(23)$,
obtained by the conformal bootstrap method, see ref. 
\cite{Simmons-Duffin:2016wlq} and references therein.
For reviews on critical phenomena see for example \cite{WiKo,Fisher74,Fisher98,PeVi}.

The concepts of critical phenomena can be extended to dynamic processes. 
For a seminal review see \cite{HoHa77}. In addition to the fundamental 
characteristics of the static universality class,
a dynamic universality class is characterized
by the type of the dynamics and whether the energy or the order parameter 
are conserved. For a detailed discussion of the classification scheme
see refs. \cite{HoHa77,FoMo06}. For a review and a book on the related 
subject of ageing see \cite{CaGa05,Malte09}.
Here we study purely dissipative relaxational dynamics without conservation of the order
parameter or the energy, which is denoted as model A in  ref. \cite{HoHa77}.

In a numerical study, the dynamics of a lattice model can be studied in various settings. 
We might consider autocorrelation times $\tau$ of systems in equilibrium
or various off equilibrium situations.  For example the system can be 
prepared in a low or high temperature state and then it is, for example,
subject to a sudden quench to the critical temperature. In the case of 
damage spreading, the system is prepared in a spatially inhomogeneous state.
The system might also be subject to a slowly varying external field.
Here, we consider equilibrium dynamics at the critical point and a 
sudden quench from a fully magnetized configuration, corresponding to zero 
temperature, to the critical one.

Roughly speaking, the autocorrelation time $\tau$ is the time 
needed to generate a statistically independent configuration in a
stochastic process at equilibrium. More precise definitions will be given
below in section \ref{Tausec}. 
In the neighborhood of a critical point the autocorrelation time
increases with increasing correlation length $\xi$.  
This phenomenon is called critical slowing down. 
The increase is governed by a power law
\begin{equation}
 \tau \simeq \xi^z \;, 
\end{equation}
where $z$ is the dynamic critical exponent. It can not be related to 
the static exponents. 
Similar to eq.~(\ref{xipower}), the power law is subject to corrections.
Below we simulate directly at the critical point, where
the linear lattice size $L$ takes over the role of the characteristic 
length scale: $\tau \simeq L^z$.  The exponent $z$ also governs 
non-equilibrium dynamics. For a detailed discussion see for example
refs. \cite{Su76,JaScSc89,Zheng98}. 

Field theoretic results for $z$ 
relevant to the problem studied here are discussed in section
9 of ref. \cite{FoMo06}. However
one should notice the considerable progress that has been achieved
recently in refs. \cite{Me15,FRG,Ad17}.

In ref. \cite{HaHoMa72} the dynamic critical exponent $z$ was computed
to two-loop order in the $\epsilon$-expansion. 
The authors express their result as
\begin{equation}
\label{splitz}
 z = 2 + c \eta \;,
\end{equation}
where $c=6 \ln(4/3)-1 = 0.72609\;...$ and $\eta$ is the static critical exponent that 
governs the decay of the two-point function at criticality. 
In ref. \cite{AnVa84} this result was extended to
three-loop order, resulting in $c=0.72609 \; (1-0.1885 \; \epsilon + ...) $,
where $d=4-\epsilon$ is the dimension of the system.
Based on the fact that the coefficient of $\epsilon$ is small, 
one might hope that  most of the difficulties in analyzing the 
series are shuffled into $\eta$ and the series of $c$ is, 
in a vague sense, well behaved.

Recently, the $\epsilon$-expansion has been extended to 
four-loop \cite{Ad17}.  Based on this result, 
in appendix \ref{analysing}  we obtain $z=2.0243$
for three dimensions,
taking also into account the accurate estimates $z=2.1665(12)$ 
\cite{NiBl96} and $z= 2.1667(5)$ \cite{NiBl00} for two dimensions
and $z = 2 + \epsilon' - \frac{1}{2} \epsilon'^2 + ...$, 
where $d=1+\epsilon'$, given in ref. \cite{Bausch81}. 

In addition to the $\epsilon$-expansion, the problem has been attacked by
a perturbative expansion in fixed dimension.  The four-loop result for three
dimensions \cite{Prud97} had been analyzed by the authors by using
a Pad{\'e} resummation, resulting in $z=2.017$, which is consistent with the 
resummation of the three-loop result \cite{Prud92}.  However, given the 
fact that a similar analysis for two dimensions gives $z=2.093$ \cite{Prud97},
one might suspect that also the result for three dimensions is too small.
This is further corroborated by our analysis given in appendix \ref{analysing}. 
For the application of different resummation schemes to the series see also
\cite{Prud06}.

Finally let us mention the estimates obtained by using functional renormalization 
group methods \cite{Me15,FRG}. In ref. \cite{Me15}, towards the end of section 
VI, the authors give their estimate $z \approx 2.025$ for the case of the 
three-dimensional Ising  universality class. In ref. \cite{FRG} numerical 
results are presented in table I of the paper.
Using three different frequency regulators, the authors get $z= 2.024$, $2.024$,
and $2.023$, respectively. Without such a regulator $z=2.032$ is obtained.
The corresponding results for two dimensions are given in table II of 
ref. \cite{FRG}.
These are $z= 2.16$, $2.15$, and $2.14$ for the three different frequency 
regulators.
Without such a regulator $z=2.28$ is obtained, 
which is quite far off from the results of 
refs. \cite{NiBl96,NiBl00}. This suggests that also in three dimensions, 
the estimates obtained with a frequency regulators should be more reliable 
than that without.

In summary, refs. \cite{Me15,FRG} and the analysis of the four-loop $\epsilon$-expansion
\cite{Ad17} now suggest 
\begin{equation}
\label{myeps}
z \approx 2.024 \;,
\end{equation}
which is somewhat larger than $z=2.021$ for the three-loop 
$\epsilon$-expansion and $z=2.017$ for the four-loop expansion in three 
dimensions fixed, which are cited in table 4 of ref. \cite{FoMo06}.

Now let us turn to  Monte Carlo (MC) simulations of lattice models.
In table \ref{literature} we summarize results for the 
exponent $z$. 
In most 
of the papers, the Ising model on the simple cubic lattice has been
studied \cite{WaLa87,WaLa91,Muenkel93,Ito93,Grassberger,Jaster99,Ito00}.
In ref. \cite{Murase}, the Ising model on the body centered 
cubic (bcc) and face centered cubic (fcc) lattice has been simulated.
Finally in ref. \cite{Collura}, similar to the present work, 
the improved Blume-Capel model on the simple cubic lattice is studied.
Improved means that the parameter of the Blume-Capel model is chosen such 
that leading corrections to scaling vanish. For the definition of the 
Blume-Capel model see section \ref{themodel} below.
In refs. \cite{WaLa87,WaLa91} equilibrium autocorrelation times are 
determined. In ref. \cite{Grassberger} damage spreading is considered.
Else short time dynamics is studied. Mostly
the simulations are started with an ordered configuration, 
corresponding to $T=0$, and a sudden quench to $T_c$ is performed.

\begin{table}
\caption{\sl \label{literature} We summarize
results for the dynamic critical exponent $z$
obtained by Monte Carlo simulations of lattice models.
Note that in refs. 
\cite{Ito93,Ito00,Murase} the exponent $\lambda=\beta/\nu z$ is computed that 
we have converted here by using $\beta/\nu = \Delta_{\sigma}= 0.5181489(10)$ in
three dimensions \cite{Simmons-Duffin:2016wlq}. In most of the cases, the 
Ising model on the simple cubic lattice is simulated. In the case of ref.
\cite{Murase}, the Ising model on the body centered cubic (bcc) and 
face centered cubic (fcc) lattice is studied.  The author of ref. 
\cite{Collura} simulates the improved Blume-Capel (BC) model on the simple 
cubic lattice. 
The temperature is denoted by $T$ and $T_c$ is the critical temperature.
}
\begin{center}
\begin{tabular}{ccll}
\hline
ref. & year & \mc{1}{c}{method} & \mc{1}{c}{$z$} \\
\hline
\cite{WaLa87}   & 1987  & equilibrium dynamic critical behavior& 2.03(4)\\
\cite{WaLa91}   & 1991  & equilibrium dynamic critical behavior& 2.03(4)\\
\cite{Muenkel93}  & 1993 & ordered, sudden quench to $T_c$  &  2.08(3) \\
\cite{Ito93}       & 1993 & ordered, sudden quench to $T_c$  &  2.073(16)     \\
\cite{Grassberger} &1995&  damage spreading &  2.032(4) \\
\cite{Jaster99}    & 1999 &  short time dynamics, various settings& 2.042(6) \\
\cite{Ito00}       & 2000 &  ordered, sudden quench to $T_c$ & 2.055(10) \\
\cite{Murase}      & 2007 &  bcc, ordered, sudden quench to $T_c$ &2.064(24)  \\
\cite{Murase}      & 2007 &  fcc, ordered, sudden quench to $T_c$ &2.056(24) \\
\cite{Collura}     & 2010 & improved BC, $T=\infty$, sudden quench to $T_c$ &  2.020(8) \\
\hline
\end{tabular}
\end{center}
\end{table}

The simulations of the Ising model give results for $z$ that are larger 
than the field theoretic ones. In particular all studies that are performed 
later than 1991 are not compatible within the quoted error bars with
eq.~(\ref{myeps}).
None of these simulations should have a principle flaw. Therefore, assuming 
the correctness of eq.~(\ref{myeps}), one might argue that the discrepancy
is due to the leading correction to scaling that is not properly taken into
account in the analysis of the data.

This was the motivation of ref. \cite{Collura} to simulate the improved 
Blume-Capel model on the simple cubic lattice instead of the Ising model.
Indeed the estimate given in ref. \cite{Collura} is fully consistent with 
the field theoretic one. 
Also here we simulate the improved Blume-Capel model, aiming at a
considerably higher accuracy than that of ref. \cite{Collura}.

Experimental results are a bit scarce. In a recent experiment 
\cite{experiment1} $\nu z \approx 1.3$ was found. Using $\nu=0.6299709(40)$, 
ref. \cite{Simmons-Duffin:2016wlq}, one gets $z \approx 2.06$.  Besides 
uncertainties in the experimental determination of data, leading corrections 
to scaling might be an issue in the analysis of the data.

In the following section we discuss the Blume-Capel model, define the 
observables that are measured and discuss briefly subleading corrections. 
Next we define the algorithms that are used. Then we discuss how the 
autocorrelation time is defined and how it is determined in the simulation.
In section \ref{NumTC}  we  study the equilibrium 
autocorrelation times of the local 
heat bath and the Metropolis algorithm on finite lattices at the critical
temperature. Then in section \ref{NERMC} we discuss our results for
a sudden quench to criticality starting from a fully magnetized configuration. 
Finally we summarize and give our conclusions. In appendix A we discuss
our implementation of the heat bath algorithm.
In appendix B we report results for the two-dimensional 
Ising model. In appendix C we analyze the four-loop $\epsilon$-expansion 
\cite{Ad17}. In appendix \ref{leadingC} we analyze leading corrections to scaling based
on simulations of the Ising model and the Blume-Capel model at $D=1.15$.

\section{The model}
\label{themodel}
The Blume-Capel model is characterized by the reduced Hamiltonian
\begin{equation}
\label{Blumeaction}
H = -\beta \sum_{<xy>} s_x s_y
  + D \sum_x s_x^2  - h \sum_x s_x \;\;  ,
\end{equation}
where the spin might assume the values $s_x \in \{-1,0,1 \}$.
$x=(x_0,x_1,x_2)$
denotes a site of the simple cubic lattice, where $x_i \in \{0,1,2,...,L_i-1\}$.
We employ periodic boundary conditions in all directions of the lattice.
Throughout we shall consider $L_0=L_1=L_2=L$ and a vanishing
external field $h=0$. 
In the limit $D \rightarrow - \infty$  the ``state"  $s=0$ is completely
suppressed, compared with $s=\pm 1$, and therefore the spin-1/2 Ising model
is recovered.
In  $d\ge 2$  dimensions the model undergoes a continuous phase transition
for $-\infty \le  D   \le D_{tri} $ at a $\beta_c(D)$.
For $D > D_{tri}$ the model undergoes a first order phase transition.
Refs. \cite{des,HeBlo98,DeBl04}  give for the three-dimensional simple cubic
lattice
$D_{tri} \approx 2.006$,  $D_{tri}\approx 2.05$ and $D_{tri} =2.0313(4)$,
respectively. It has been demonstrated numerically that on the line of 
second order phase transitions, there is a point $(D^*,\beta_c(D^*))$, 
where the amplitude of 
the leading correction to scaling vanishes, see ref. \cite{myBC} and 
references therein. Following ref. \cite{myBC}
\begin{equation}
 D^* = 0.656(20)
\end{equation}
and
\begin{equation}
\beta_c(D=0.655)=0.387721735(25) \;\;.
\end{equation}
Here we simulated at $(D,\beta)=(0.655,0.387721735)$. 
At $D=0.655$ leading corrections to scaling should be at least by
a factor of 30 smaller than in the spin-1/2 Ising model on the simple cubic 
lattice.

In ref. \cite{myBC} we obtained $\nu=0.63002(10)$, $\eta=0.03627(10)$ and 
$\omega=0.832(6)$, which were nicely confirmed by the conformal bootstrap 
method. Note that also the accurate estimates of surface critical exponents
for the ordinary and special surface universality classes that we 
obtained by simulating the improved Blume-Capel model in refs. 
\cite{MHordinary,MHspecial} were confirmed by using the conformal 
bootstrap method \cite{Gliozzi}.

\subsection{The observables}
We focus on the magnetization 
\begin{equation}
m = \frac{1}{L^3} \sum_x s_x
\end{equation}
and the estimator of the magnetic susceptibility  
\begin{equation}
\chi  \equiv  \frac{1}{L^3}
\Big(\sum_x s_x \Big)^2 \;\;
\end{equation}
for a vanishing expectation of the magnetization. The Binder 
cumulant 
\begin{equation}
U_4 = \frac{\langle m^4 \rangle}{\langle m^2 \rangle^2}
\end{equation}
is the prototype of a dimensionless quantity and is 
well suited to detect leading corrections to scaling.

Furthermore we measured
\begin{equation}
\label{energy}
 E = \frac{1}{L^3}  \sum_{<xy>}  s_x  s_y \;\;,
\end{equation}
which is proportional to the energy density.

\subsection{Subleading corrections to scaling}
Below we analyze the behavior of the magnetic susceptibility $\chi$ 
and the integrated autocorrelation time $\tau_{int,\chi}$ of $\chi$
at the critical temperature on finite lattices of the linear size $L$.
For the definition of $\tau_{int,\chi}$ see section \ref{Tausec} below.
In the case of the magnetic susceptibility we expect
\begin{equation}
\label{chifinite} 
 \chi = a(D) L^{2-\eta} \;\left(1 + b_1(D) L^{-\omega} + b_2 b_1^2(D) L^{-2 \omega}  + ... 
+ c(D)  L^{-\omega_2} + ... \right) \; + B(D) \; ,
\end{equation}
where $\omega_2$ is the exponent of the subleading correction and $B(D)$ is the 
analytic background. The argument $D$ is the parameter of the Blume-Capel
model, eq.~(\ref{Blumeaction}). Since $b_1(0.655) \approx 0$, in our data 
for $D=0.655$, 
subleading corrections are actually the numerically dominating ones.
Eq.~(\ref{chifinite}) can be obtained for example by taking the second derivative with 
respect to the external field of both sides in eq.~(2.14) of ref. \cite{PeVi}.

In ref. \cite{myBC} we assumed  $\omega_2=1.67(11)$,  obtained 
by using the scaling field method \cite{NewmanRiedel}.  Note however that
for even, rotationally invariant perturbations to the fixed point,
the authors of ref. \cite{Litim04} find, by using the functional 
renormalization group method, clearly larger values. In table 3 of 
ref. \cite{Litim04} estimates $\omega_2=2.838$ up to $3.6845$, depending on 
the cutoff scheme that is used, are given. In table 2 of ref. 
\cite{Simmons-Duffin:2016wlq} the accurate estimate 
$\Delta_{\epsilon''}= 6.8959(43)$, corresponding to $\omega_2=3.8959(43)$
is given.  We conclude that $\omega_2=1.67(11)$ is an artifact of the 
scaling field method. 

One should notice that the magnetic susceptibility at the critical point 
on a finite lattice is affected by the breaking of the rotational symmetry 
by the simple cubic lattice. The analogous fact has been demonstrated very 
clearly
for the two-dimensional Ising model on the square lattice \cite{our2D}.
See in particular section 6, where data obtained by using the numerical 
transfer matrix method are analyzed. In two dimensions the corresponding
correction exponent is $\omega_{NR}=2$.  It is interesting to note that
the correction is related to the interplay between the torus geometry
and the square lattice. For temperatures different from the critical
one, in the thermodynamic limit, the correction is absent in the 
magnetic susceptibility, see ref. \cite{our2D}
and references therein.

In the case of the 
three-dimensional Ising universality class,  $\omega_{NR} = 2.0208(12)$, see
table 1 of ref. \cite{Campostrini:2002cf}, or more recently 
$\omega_{NR}=2.022665(28)$ obtained from the
scaling dimension of the even operator with spin $l=4$, given 
in table 2 of ref. \cite{Simmons-Duffin:2016wlq}.  
Note that $\omega_{NR}=2.022665(28)$ is clearly smaller than 
$\omega_2=3.8959(43)$. 

For a brief discussion on the breaking of the rotational symmetry by the 
lattice and corrections see also section 1.6.4 of ref. \cite{PeVi}.

The analytic background $B(D)$ can be viewed as a correction with the 
correction exponent $2-\eta$. Given the accuracy of our data it is 
useless to put two correction terms with almost degenerate exponents
into an ansatz.
Instead we use a single term proportional to $L^{-\epsilon}$ that effectively 
takes into account both corrections. Mostly we set $\epsilon=2$. 
The exponent is denoted by $\epsilon$ 
to indicate that it is an effective correction exponent used in the analysis of the
data.

\section{The algorithms}
\label{algor}
We perform two different types of simulations. First we studied the 
equilibrium behavior at the critical point for finite lattices. 
In this case we used a hybrid of the single cluster algorithm \cite{Wolff} 
and  local updates to efficiently equilibrate the system.  In the 
hybrid update, sweeps using the local algorithm alternate with a certain 
number of single cluster updates.  Our measurements are organized in bins.
These bins are separated by hybrid updates. While measuring only local 
updates are performed.
In a second set of simulations 
we started from fully magnetized configurations corresponding to 
zero temperature. In a sudden quench, the temperature is set to the critical
value. Here, of course, only local updates are used.

As local update we used either the heat bath algorithm or the 
particular Metropolis algorithm discussed in section IV of ref. \cite{myBC}.
The simulation programs are written in C. 
The heat bath algorithm used in the first stage of the study, discussed
in section \ref{comparison}, 
is implemented in a more or less straightforward way, storing the spins
as \verb+char+ variables. The simulations in sections \ref{main} and
\ref{HeatNER}, 
which were performed at a later stage, where performed 
by using a version of the program that is partially parallelized by using 
\verb+SSE2+ intrinsics. Furthermore, the random number is used fourfold, as
discussed in section \ref{HeatNER} below. Details are discussed in 
appendix \ref{SSE2program}. As random number generator, we have mostly used 
the SIMD-oriented Fast Mersenne
Twister algorithm \cite{twister}. Equilibrium simulations for a few lattice
sizes were partially performed by using the WELL random number generator
\cite{well}, giving consistent results.

In the case of the Metropolis algorithm, we are using  multispin coding and 
64 systems are simulated in parallel. Here $64$ is the number of bits contained 
in a long integer variable. As discussed in ref. \cite{myBC}, we were not able
to take advantage of the multispin coding when using the cluster algorithm. 
Hence we update the 64 systems one by one when performing the cluster update. 
As random number generator, we have used the SIMD-oriented Fast Mersenne
Twister algorithm \cite{twister}. 

In section \ref{comparison} we compare various orderings of the local
update scheme.  In the major simulations, we divide the
lattice in checkerboard fashion and update the sublattices alternately.

Our simulations were performed on various PCs and servers. 
In addition to the parallelization discussed above, several instances 
of the program were run with different seeds of the random number generator.
As a typical example let us quote the times needed on a single
core of an Intel(R) Xeon(R) CPU E3-1225 v3 running at 3.20GHz.
In the case of the heat bath algorithm
we need about 5 ns for the update of a single site. The time needed 
for the measurement of the energy and the magnetization is 1.8 ns for 
one site. The parallel version of the program, with a fourfold reuse of the 
random number, takes 1.8 ns for the update of a single site. The measurement
of the magnetization takes 0.1 ns per site.
In the case of the Metropolis algorithm, implemented by using 
multispin coding, 0.9 ns 
are needed for the update of a single site. The measurement of 
the energy and the magnetization, implemented by using multispin coding, 
takes about 0.3 ns per site.  

\section{The autocorrelation time}
\label{Tausec}
In the simulations at equilibrium we determined the integrated autocorrelation
time. Let us briefly recall the basic definitions.
Let us consider a generic estimator $A$. The autocorrelation function of $A$
is defined by 
\begin{equation}
 \rho_A(t) = \frac{\langle A_{i} A_{i+t} \rangle - \langle A \rangle^2}
                {\langle A^2 \rangle - \langle A \rangle^2} \;\;,
\end{equation}
where we average over the times $i$. 

If the Markov process fulfills detailed balance the eigenvalues of the 
transition matrix are real and
hence
\begin{equation}
\label{expdecay} 
 \rho_A(t) = \sum_{\alpha} a_{A,\alpha}  \exp(- t/\tau_{exp,\alpha})  \;\;. 
\end{equation}
Note that even if the local update fulfills detailed balance, as it is the 
case for the heat bath and Metropolis algorithm used here, the composite 
update, consisting of an ordered sweep over the lattice, does not. 
However, often 
one still finds that eq.~(\ref{expdecay}) is a good approximation of the 
behavior of $\rho_A$. For a discussion see for example \cite{MaSo88,So89,Wolfftau}. 

Our goal is to find a quantity that is proportional to the 
exponential autocorrelation time
$\tau_{exp}=\mbox{max}_{\alpha} \tau_{exp,\alpha}$ and that can be determined 
in the simulation with 
small statistical and systematical errors. 

Our starting point is the  integrated autocorrelation time 
\begin{equation}
\tau_{int,A}= \frac{1}{2} + \sum_{t=1}^{\infty} \rho_A(t) \;.
\end{equation}
In a numerical study the summation has to be truncated. In practice the 
upper bound is taken, selfconsistently, as a few times $\tau_{int,A}$. 
See for example \cite{MaSo88,So89,Wolfftau}. 
Since we intend to reduce effects of the 
truncation, we continued the sum, assuming a single exponential decay:
\begin{equation}
\tau_{int,A}= \frac{1}{2} + \sum_{t=1}^{t_{max}} \rho(t) 
 + \sum_{t=t_{max}+1}^{\infty} \tilde \rho(t) \;\;,
\end{equation}
with
\begin{equation}
 \tilde \rho(t)  = a(t_{max}) \exp(-t/\tau_{eff}(t_{max})) \;\;,
\end{equation}
where 
\begin{equation}
\tau_{eff}(t) = - 1/\ln[\rho_A(t+1)/\rho_A(t)]
\end{equation} 
and 
\begin{equation}
a(t)  =  \rho(t) \exp(t/\tau_{eff}(t))  \;\;.
\end{equation}
We determined the  autocorrelation function of the energy density, the 
magnetization and the magnetic susceptibility. Preliminary studies 
have shown that the scaling of the integrated autocorrelation time of these
three quantities with the linear lattice size $L$ is consistent. Also 
plotting $\tau_{eff}(t)/\tau_{int}$ as a function of $(t+1/2)/\tau_{int}$
we find a collapse of the data for different lattice sizes.

To keep the study tractable, we focus on the integrated autocorrelation 
time $\tau_{int,\chi}$ of the magnetic susceptibility in the following.
Throughout we take $t_{max} \approx 3 \tau_{int,\chi}$.

In our simulations, the autocorrelation functions are computed in the 
following way. We consider distances $t$ up to  $t_{MAX}>t_{max}$.
The simulations are organized in bins of the size $(n_t + 1) \; t_{MAX}$, 
where $t_{MAX} = 2 L^2$ throughout.  Here we make use of the fact that 
$z \approx 2$. 
Then 
\begin{equation}
 \overline{A} = \frac{1}{n_t \; t_{MAX}} \sum_{i=1}^{n_t\; t_{MAX}} A_i \;,
\end{equation}
\begin{equation}
 \overline{A^2} = \frac{1}{n_t \; t_{MAX}} \sum_{i=1}^{n_t \; t_{MAX}} A_i^2  
\end{equation}
and
\begin{equation}
\overline{A_i A_{i+t}}=\frac{1}{n_t \; t_{MAX}} \sum_{i=1}^{n_t \;t_{MAX}} A_i
A_{i+t} \;\;.
\end{equation}
For each bin, these averages are stored in a file for the subsequent 
analysis. Statistical errors are computed by using the jackknife method.

\section{Equilibrium autocorrelation times at the critical temperature}
\label{NumTC}
Since we discuss only the integrated autocorrelation time of the 
magnetic susceptibility in the following, we mostly drop for simplicity the 
subscript of $\tau_{int,\chi}$. In a preliminary study
we compared the autocorrelation times of the heat bath algorithm 
using different orders of the local update and that of our Metropolis algorithm
with checkerboard decomposition.  To this end we simulated a number of linear
lattice sizes up to $L=28$.  We conclude that the difference in the behavior
of the autocorrelation times is compatible with an overall factor and 
corrections that decay like $L^{-2}$.

Next we performed simulations with an increased statistics and larger lattice
sizes using the heat bath algorithm and our Metropolis algorithm, in both 
cases using a checkerboard decomposition. 

In order to check the effect of leading corrections to scaling we 
have simulated the Ising model and the Blume-Capel model at $D=1.15$
by using the heat bath algorithm using a checkerboard decomposition for 
linear lattice sizes up to $L=24$. The results are discussed in appendix 
\ref{leadingC}. 

\subsection{Comparing various local update schemes at the critical point}
\label{comparison}
As a comparison of the performance, and check whether different local 
updates result in the same exponent $z$, we did run simulations
for lattice sizes $L=8$, $10$, $12$, $14$, $16$, $20$, $24$, and $28$ at $D=0.655$ 
and $\beta=0.387721735$.
We performed local heat bath (HB) updates, visiting the sites of the lattice 
in different order. In the first case, denoted by $C$, we divide the lattice 
in checkerboard fashion. The two sublattices are updated alternately.
Running through the lattice in typewriter fashion is denoted by $T$. 
Finally, the site that is updated is selected randomly.  This
is denoted by $R$. A unit of time has passed, when $L_0 L_1 L_2$ sites 
have been updated. The Metropolis (M) algorithm is only simulated with
checkerboard decomposition.

We fitted ratios of integrated autocorrelation times of the magnetic 
susceptibility with the ansatz
\begin{equation}
\label{ratioansatz}
 \frac{\tau_{A_1}(L)}{\tau_{A_2}(L)} = r \; (1 + a L^{-\epsilon})  \;,
\end{equation}
where $r$ and $a$ are free parameters.  Here, $A_1$ and $A_2$ denote the two different 
algorithms that have been used.  We fix the correction exponent $\epsilon=2$.
In table \ref{compare} we summarize our results. In these fits, 
all lattice sizes $8 \le L \le 28$ are taken into account.

\begin{table}
\caption{\sl \label{compare} We give the results for the comparison 
of different  local update algorithms $A_1$ and $A_2$. $r$ and $a$ are
the free parameters of the ansatz~(\ref{ratioansatz}). 
}
\begin{center}
\begin{tabular}{ccccc}
\hline
 $A_1$ & $A_2$  &  $r$  & $a$ & $\chi^2$/d.o.f. \\
\hline
$(HB,R)$ & $(HB,C)$ & 1.98990(43) & -0.282(29) & 1.06 \\  
$(HB,T)$ & $(HB,C)$ & 0.99985(20) & -0.106(21) & 0.32 \\  
$(M,C)$  & $(HB,C)$ & 1.33136(19) & -2.100(12) & 0.98 \\  
\hline
\end{tabular}
\end{center}
\end{table}
We conclude that the different local update schemes are indeed
characterized by the same dynamic critical exponent $z$.
Corrections in the ratios of autocorrelation times 
vanish quickly, consistent with a behavior $\propto L^{-2}$.

\subsection{Heat bath and Metropolis algorithm with checkerboard decomposition}
\label{main}
We simulated a large number of linear lattice sizes
up to $L=56$ and $72$ using the Metropolis and the heat bath algorithm,
respectively. In total these simulations took the equivalent of 
of about 2.8 and 5.6 years, respectively, of CPU time on 
one core of a Intel(R) Xeon(R) CPU E3-1225 v3 CPU. 

To give the reader an 
impression of the accuracy of the numbers, we quote $\chi= 2558.23(24)$ and
$\tau_{int,\chi} = 964.44(30)$ for $L=56$ obtained from the simulations with 
the Metropolis algorithm. The simulation for $L=56$ consists of 388 bins.
Each bin contains 64 replicas that were simulated in parallel performing
$(1001 \times 2 \times 56^2)$ full lattice updates for each replica.
 Using the heat bath algorithm we get 
$\chi=4189.1(1.0)$ and $\tau_{int,\chi} = 1206.1(1.1)$ for $L=72$.
The simulation for $L=72$  consists of 1359 bins containing 
16 replicas that were simulated in parallel performing
$(101 \times 2 \times 72^2)$ full lattice updates of each replica and bin.

As a benchmark we first analyze the behavior of the
magnetic susceptibility at the critical point. The result for the 
critical exponent $\eta$ can be compared with the accurate estimate 
obtained by the conformal bootstrap method.  It follows the analysis
of the autocorrelation times obtained for the Metropolis and heat bath
algorithms.

\subsubsection{The magnetic susceptibility}
First we checked that
the results obtained for $D=0.655$ by using the Metropolis and the 
heat bath algorithm are consistent. Below we analyze the merged results.
Assuming that the amplitude of leading corrections to scaling vanishes, 
we fitted the data with the ans\"atze
\begin{equation}
\label{chi1}
\chi= a L^{2-\eta}
\end{equation}
and 
\begin{equation}
\label{chi2}
\chi = a L^{2-\eta} \; (1 + b L^{-\epsilon})  \; ,
\end{equation}
where we have taken $\epsilon=2$. We checked that replacing 
$\epsilon=2$ by $\epsilon=2-\eta=1.9637022$ or 
$\epsilon=\omega_{NR}=2.022665$ changes the estimate of $\eta$ by little. 
Of course, still we can not exclude that
these two corrections have amplitudes with opposite sign and cancel
to a large extend in the range of lattice sizes considered here.
In the fits all data for lattice sizes $L \ge L_{min}$ are taken into account.

\begin{figure}
\begin{center}
\includegraphics[width=14.5cm]{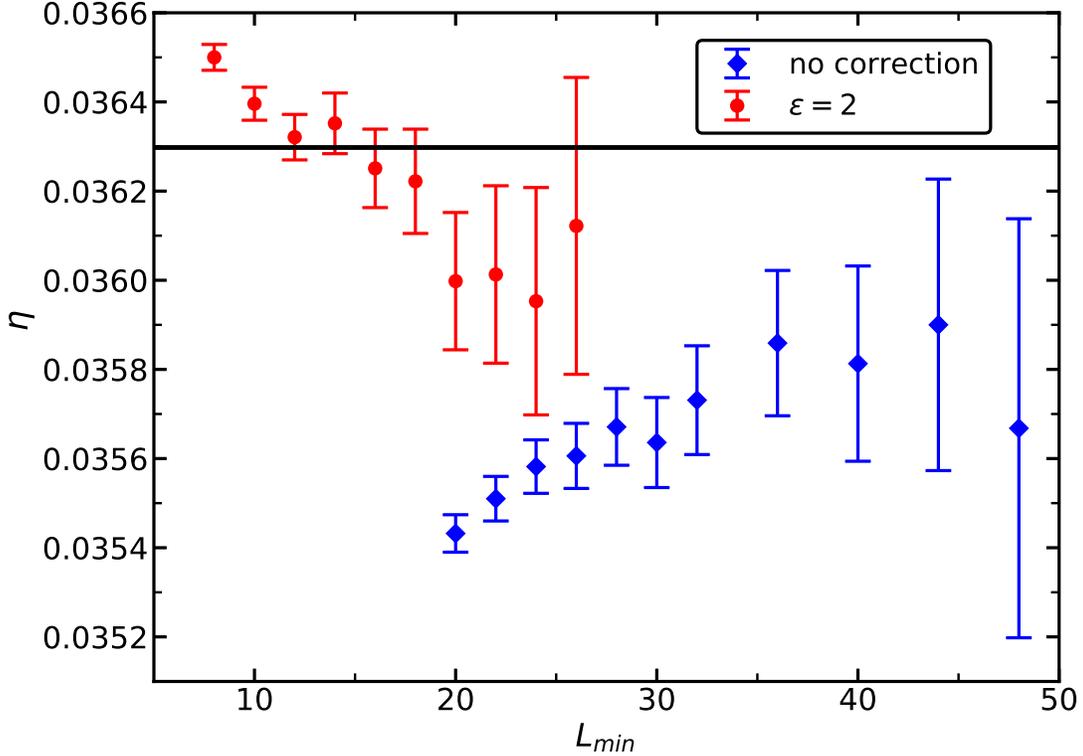}
\caption{\label{etaplot}
Results for the critical exponent $\eta$ obtained by fitting our 
numerical estimates of the magnetic susceptibility $\chi$ at $D=0.655$
and $\beta=0.387721735$ by using the ans\"atze~(\ref{chi1},\ref{chi2}).
The solid line indicates the result obtained from conformal 
bootstrap $\eta=0.0362978(20)$, ref. \cite{Simmons-Duffin:2016wlq}.
All data with $L \ge L_{min}$ are taken into account in the fit.
}
\end{center}
\end{figure}

In figure \ref{etaplot} we plot the results for $\eta$ obtained from these fits
as a function of $L_{min}$.
In the case of the ansatz~(\ref{chi2}) we find $\chi^2/$d.o.f.$=2.31$,
$1.32$, $1.12$, $1.16$, $1.02$, $1.09$, $0.75$, $0.82$, $0.89$, and $0.92$
for $L_{min}=8$, $10$, $12$, $14$, $16$, $18$, $20$, $22$, $24$ and 
$26$, respectively.  
The large value of $\chi^2/$d.o.f. for $L_{min}=8$ and $10$ indicates that 
corrections, for example $\propto L^{-\omega_2}$ with
$\omega_2=3.8959(43)$, that are not taken into account in the
ansatz~(\ref{chi2}) give, at least for $L=8$ and $10$, contributions to
$\chi$ that are larger than the statistical error of our estimate.
Starting from $L_{min}=12$ the ansatz~(\ref{chi2}) is not ruled
out by the $\chi^2/$d.o.f., yet corrections that are not contained in the 
ansatz might still cause a systematic error of the estimate of the exponent 
$\eta$.
The estimates
of the correction amplitude are $b=-0.4248(25)$, $-0.410(4)$, $-0.393(9)$, 
$-0.402(15)$, $-0.368(24)$, $-0.356(40)$, $-0.244(64)$, $-0.253(10)$, 
$-0.21(14)$, and $-0.34(21)$ for $L_{min}=8$, $10$, $12$, $14$, ..., and $26$,  
respectively. With increasing $L_{min}$ the statistical error of $b$ rapidly
increases. For $L_{min}=26$, the statistical error of $b$ is almost as large
as its absolute value.  Therefore fitting the data for even larger $L_{min}$ 
with the ansatz~(\ref{chi2}) is useless.
In the case of 
the ansatz~(\ref{chi1}) we find $\chi^2/$d.o.f.$=1.82$, $1.32$, 
$1.01$, $1.08$, $0.96$, $1.03$, $0.90$, $0.82$, $0.96$, $1.17$ and $1.40$
for $L_{min}=20$, $22$, $24$, $26$, $28$, $30$, $32$, $36$, $40$, $44$, and
$48$, respectively.

While our estimate of $\eta$ using the 
ansatz~(\ref{chi2}) and $L_{min}=12$ is compatible with the 
conformal bootstrap result,  the estimate from the ansatz~(\ref{chi1}) 
and $L_{min}=24$ is by 9 times the error bar too small compared with the 
conformal bootstrap.  This is a nice reminder of the fact that 
$\chi^2/$d.o.f. $\approx 1$ does not guarantee that the effects of 
corrections that are not taken into account in the ansatz are small.
One might try to estimate these systematic effects by comparing the results obtained 
from different ans\"atze. In the present case, the difference 
between the estimate of $\eta$ obtained from the ansatz~(\ref{chi2}) 
for $L_{min}=12$ and the ansatz~(\ref{chi1}) for $L_{min}=24$ could
serve this purpose. Since the estimate of $\eta$ 
obtained from the ansatz~(\ref{chi2}) for $L_{min}=12$ is fully consistent 
with the conformal bootstrap we will give in the analysis of the 
autocorrelation time below some preference to the ansatz that contains
a correction term. 

Finally we check the possible effect of residual leading corrections
to scaling at $D=0.655$. In ref. \cite{myBC}, we conclude that
compared with the Ising model on the simple cubic lattice, leading 
corrections to scaling are suppressed at least by a factor $1/30$. 
Based on that we have generated synthetic data by multiplying our data for 
$D=0.655$ by the factor $(1 \pm [0.22/30] L^{-\omega})$, where the
coefficient $0.22$ stems from the analysis for the Ising model discussed
in appendix \ref{leadingC}. 
Using these synthetic data we have repeated the fits using the
ans\"atze~(\ref{chi1},\ref{chi2}).  In the case of the ansatz~(\ref{chi2})
for $L_{min}=16$ we find that the estimate of $\eta$ changes by
$\pm 0.00023$. For the ansatz~(\ref{chi1}) and $L_{min}=36$ we
find that the estimate of $\eta$ changes by $\pm 0.00026$. 

Finally we consider the quantity  $\chi_{imp}=U_4^x \chi$.  
The construction of such quantities is for example discussed in ref.
\cite{ourdilute}.
The exponent
$x$ is taken such that leading corrections to scaling in $U_4^x$ and 
$\chi$ cancel. Analyzing our data for the Ising model and the 
Blume-Capel model at $D=1.15$  we find $x=-1.4$, where the error is small
enough to ensure a reduction of the amplitude of 
the leading correction to scaling 
by one order of magnitude.  Fitting $\chi_{imp}$ with the 
ansatz~(\ref{chi2}) we find $\eta=0.03625(17)$ for $L_{min}=16$ and 
with the ansatz~(\ref{chi1}) we find $\eta=0.03588(30)$ for $L_{min}=36$.  
Note that 
in particular the result obtained with the ansatz~(\ref{chi2}) is in very
good agreement with the conformal bootstrap.

\subsubsection{The scaling behavior of the autocorrelation time}

First we fitted the ratio $\tau_{M,C}/\tau_{HB,C}$ using the 
ansatz~(\ref{ratioansatz}), where now the exponent $\epsilon$ 
is a free parameter. We get $\chi^2/$d.o.f. $=0.97$ taking all lattice
sizes into account. We get $\epsilon=2.097(23)$, $ 2.167(44)$, 
$2.135(80)$, and $2.04(13)$, for $L_{min}=8$, $10$, $12$, and $14$, 
respectively, where all linear lattice sizes $L \ge L_{min}$ are taken 
into account.

We have fitted our results using the basic ans\"atze
\begin{equation}
\label{simpleA}
 \tau = a_A L^z 
\end{equation}
and
\begin{equation}
\label{correctionA}
\tau = a_A L^z \;\; (1 + c_A L^{-\epsilon}) \;,
\end{equation}
where $a_A$, $c_A$ and $z$ are the free parameters.
The subscript $A$ denotes the algorithm that is used.
We have fixed the 
correction exponent $\epsilon = 2$. Replacing the $2$ by $2-\eta$ or 
$\omega_{NR}$ has only little effect on the results for $z$.

\begin{figure}
\begin{center}
\includegraphics[width=14.5cm]{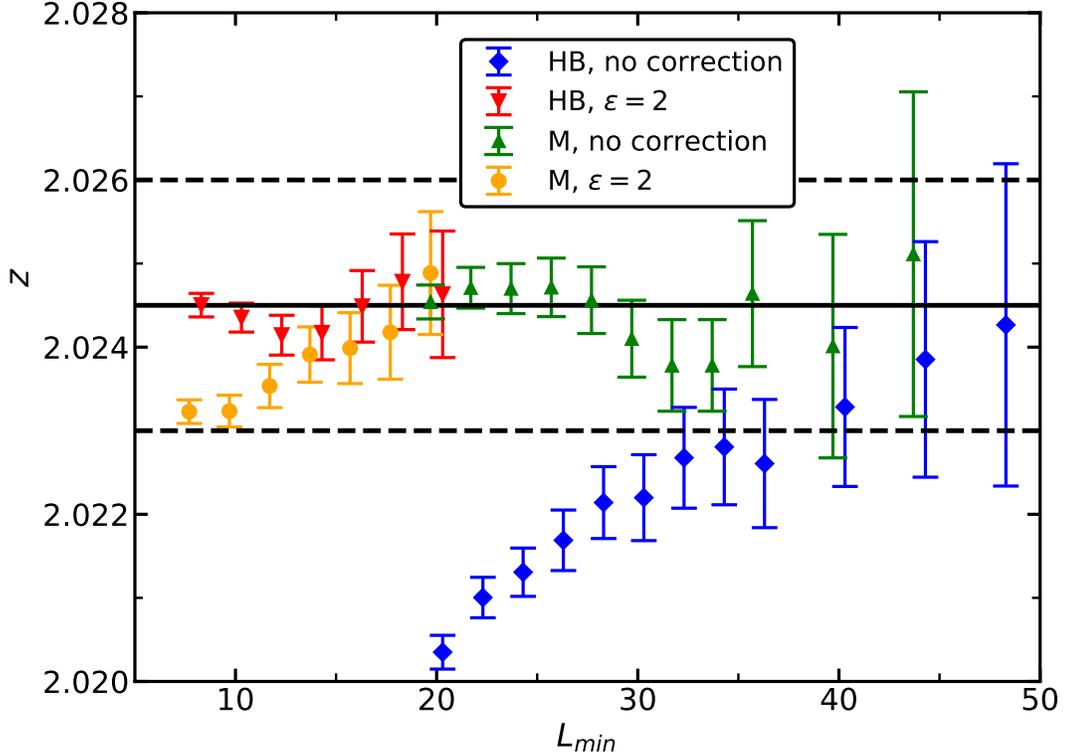}
\caption{\label{zplot}
Results for the dynamic critical exponent $z$ obtained from fitting our
numerical results for the integrated autocorrelation time of the magnetic 
susceptibility $\tau_{int,\chi}$ at $D=0.655$
and $\beta=0.387721735$ by using the ans\"atze~(\ref{simpleA},\ref{correctionA}).
All data with $L \ge L_{min}$ are taken into account in the fit.
In the caption, the Metropolis and the heat bath algorithm are indicated  
by M and HB, respectively. For better readability we have slightly shifted the 
values of $L_{min}$.  The solid line gives the central value of our estimate 
$z=2.0245(15)$, while the dashed lines indicate the error.
}
\end{center}
\end{figure}

In a first series of fits we analyzed the data for the heat bath and the 
Metropolis updates separately. The results for the exponent $z$ are 
shown in figure \ref{zplot}. In the case of the heat bath algorithm and 
the ansatz~(\ref{correctionA}) we find that  $\chi^2/$d.o.f.$<1$ starting 
from $L_{min}=8$. In the case of the Metropolis algorithm and 
the ansatz~(\ref{correctionA}) we get $\chi^2/$d.o.f.$=1.46$, $1.35$, and
$1.20$ for $L_{min}=10$, $12$, and $14$. For larger $L_{min}$ it fluctuates 
at this level. In the case of the Metropolis algorithm and the 
ansatz~(\ref{simpleA})  we have $\chi^2/$d.o.f.$=1.20$ for $L_{min}=20$.
This is related with the fact that fits with the ansatz~(\ref{correctionA}) 
give small values for the correction amplitude $c_{M}$.  In the case of 
the  heat bath algorithm and the ansatz~(\ref{simpleA}) we find 
$\chi^2/$d.o.f.$=2.98$ for $L_{min}=20$, dropping below one at $L_{min}=28$.

Next we check the possible effect of residual leading corrections to scaling.
To this end, we multiply our data for $\tau$ of the 
heat bath algorithm with $1\pm [0.43/30] L^{-\omega}$. For $L_{min}=16$ and the 
ansatz~(\ref{correctionA}) the estimate of $z$ changes by $\pm 0.00045$.
For $L_{min}=36$ and the ansatz~(\ref{simpleA}) it changes by $\pm 0.00049$.
To see what we would get for the Ising model, we multiplied the data 
with $1 - 0.43 L^{-\omega}$. Fitting with the ansatz~(\ref{simpleA}) 
we get $\chi^2/$d.o.f.$<1$ starting from $L_{min}=30$.  For example
for $L_{min}=32$ we get $z=2.0385(6)$. Note that 
$(2.0385-2.024)/0.0006\approx 24$. 

We did not simulate the Ising model or the Blume-Capel model with our 
Metropolis algorithm, since it seems to be a safe guess that the 
effect of leading corrections is much the same as for the heat bath algorithm. 

We also performed a joint fit of the Metropolis and the heat bath data 
using the ansatz~(\ref{correctionA}), where $a_M$, $c_M$, $a_{HB}$, $c_{HB}$,
and $z$ are the free parameters. For example we get $z=2.02424(30)$
for $L_{min}=16$.  Note that $\chi^2/$d.o.f.$<1$ already for $L_{min}=12$. 

Finally we fitted the improved autocorrelation time $\tau_{imp}=U_4^x \tau$,
where $x=-3.1$ for the heat bath algorithm. Here we find $z=2.02462(46)$ for 
$L_{min}=16$.  

Focusing on the fits with the ansatz~(\ref{correctionA}) and 
$L_{min}=14$, $16$ and $18$ we arrive at the estimate
\begin{equation}
 z = 2.0245(15) \;\;.
\end{equation}
The error bar covers all the fits that we performed with the 
ansatz~(\ref{correctionA}) and $L_{min}=14$, $16$ and $18$.  Also
possible effects of residual leading corrections to scaling are taken 
into account. The error bar also covers the fits of the autocorrelation times
for the Metropolis algorithm using the ansatz~(\ref{simpleA}). In the 
case of the heat bath algorithm and the ansatz~(\ref{simpleA}) at least the 
central values are covered for $L_{min} \ge 40$.  Completely covering also
the error bars of these fits, in particular in the light of the 
results for the exponent $\eta$ in the section above, seems to be too 
pessimistic.  Since $\beta_c$ was determined in \cite{myBC} using larger
lattices and higher statistics than here, it seems safe to ignore the 
error induced by the uncertainty of the estimate of $\beta_c$.  

\section{Sudden quench from $T=0$ to criticality}
\label{NERMC}
We have simulated 
the Blume-Capel model at $D=0.655$ by using our Metropolis algorithm 
and the heat bath algorithm both with checkerboard 
ordering. At time $t=0$, 
we start with a fully magnetized configuration corresponding to 
zero temperature and perform a sudden quench to $\beta=0.387721735$, 
which is our estimate of the inverse of the critical temperature \cite{myBC}.
Updating all sites of the lattice once is taken as unit of time.
In the analysis we focus for simplicity on the magnetization.
In the thermodynamic limit, it behaves as 
\begin{equation}
\label{NERmag}
m(t) = a (t-t_0)^{-\lambda_m} \;,
\end{equation}
where $\lambda_m=\beta/\nu z$.  See eq.~(2) of ref. \cite{Ito93} and references therein. 
 Note that $\beta/\nu = \Delta_{\sigma}= 0.5181489(10)$ in three
dimensions \cite{Simmons-Duffin:2016wlq}. 
Eq.~(\ref{NERmag}) is subject to leading corrections of the equilibrium 
universality class. Since we simulate an improved model, we ignore these 
corrections in our analysis.
We only take explicitly into account analytic corrections that are
expressed by $t_0$.

\subsection{Simulations using the Metropolis algorithm}
\label{MetroNER}
Most of our simulations were performed
using lattices of the linear size $L=300$. As a check of finite size effects, 
we performed simulation with $L=50$ and $100$ in addition.  In the case of
$L=50$ and $100$ we performed $2000 \times 64$ runs and for $L=300$ we performed 
$4000 \times 64$ runs. For $L=50$ we did run up to $t=1000$ and for 
$L=100$ and $300$  up to $t=4000$.
Statistical errors are computed by using the jackknife method.  
Using the multispin coding technique, 64 runs are performed in parallel,
partially sharing the same pseudo random number stream, possibly causing
a statistical correlation. Therefore these runs are always put in the same
jackknife bin, not to corrupt the estimate of the statistical error.

\begin{figure}
\begin{center}
\includegraphics[width=14.5cm]{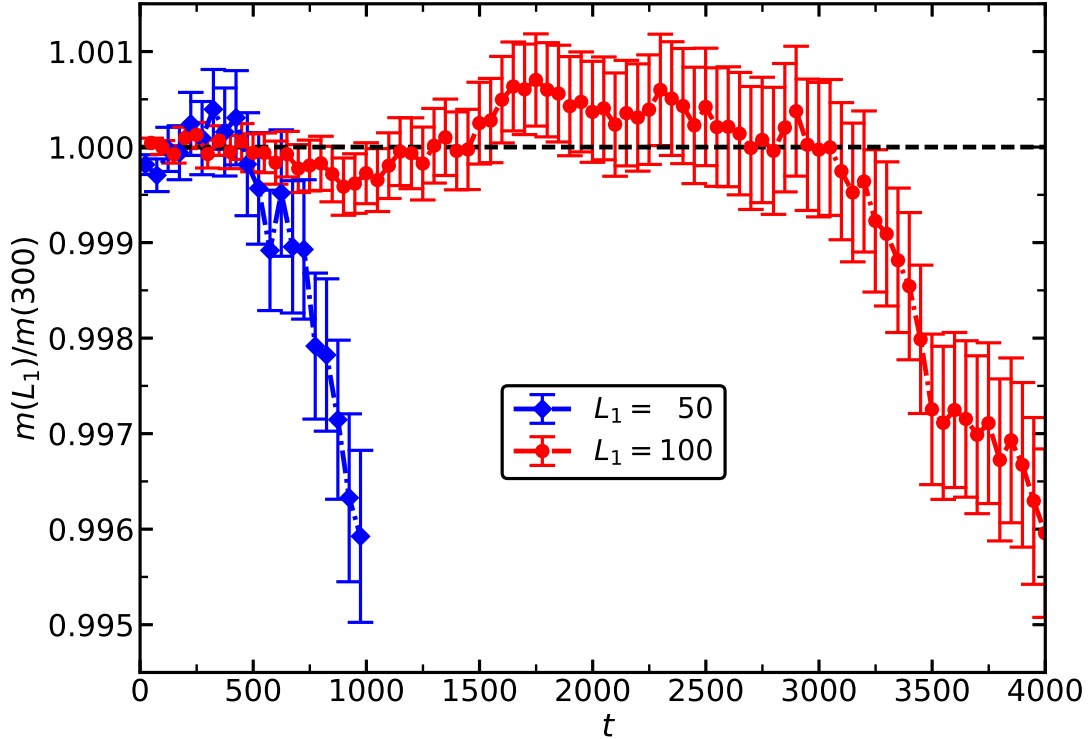}
\caption{\label{rm50x100}
Simulations with the Metropolis algorithm.
We plot the ratios $m_{L_1}(t)/m_{L_2}(t)$ for 
$L_1=50$ and $100$ and $L_2=300$ as a function of $t$.  For the readability 
of the figure we only give a fraction of the $t$ values.
}
\end{center}
\end{figure}

In figure \ref{rm50x100} we plot ratios of the magnetization as a 
function of the  Monte Carlo time $t$. We find that for $L=50$ the deviation 
from $L=300$ reaches a $3 \sigma$ level for $t \gtrapprox 840$.
For $L=100$ this is the case for $t \gtrapprox 3500$. In both cases 
we regard the magnetization for $L=300$ as approximation of the 
thermodynamic limit. From scaling we expect
that the point of deviation from the thermodynamic limit by a certain 
fraction behaves as $t \propto L^z$. Therefore
we conclude that for $L=300$ up to $t=4000$ deviations from the thermodynamic
limit can be safely ignored at the level of our statistics.
In the following only data obtained for $L=300$ are considered.  In total, 
the simulations for $L=300$ took the equivalent of about 440 days on 
a single core of a Intel(R) Xeon(R) CPU E3-1225 v3 CPU.

By construction the data for the magnetization at different values of $t$ 
are correlated. We tried to avoid fitting a large data set with correlations 
and keep the analysis simple.
Our starting point is an effective exponent given by 
\begin{equation}
\label{zeff}
z_{m,eff,t_0}(t) = - \Delta_{\sigma}
                 \frac{\ln\left[(2 t -t_0)/(t -t_0)\right]}
                    {\ln\left[m(2 t)/m(t)\right]} \;,
\end{equation}
where $t_0$ remains a free parameter.

In a first step of the analysis we fix $t_0$ by requiring that 
$z_{m,eff,t_0}(t)$ has a minimal variance in the interval $t_1 \le t < t_2$: 
The average of $z_{m,eff,t_0}$ in the interval is denoted by
\begin{equation}
 \bar z_{m,eff,t_0}(t_1,t_2) =\frac{1}{t_2-t_1} \sum_{t=t_1}^{t_2-1}
    z_{m,eff,t_0}(t)  \;\;.
\end{equation}
Then we minimize 
\begin{equation}
\label{mini}
\mbox{var}(z,t_0,t_1,t_2) = \sum_{t=t_1}^{t_2-1} 
\left[z_{m,eff,t_0}(t) -\bar z_{m,eff,t_0}(t_1,t_2) \right]^2
\end{equation}
with respect to $t_0$. The results of this analysis for $t_2=2 t_1$ and 
various values of $t_1$ are given in table \ref{fitz}.  With increasing 
$t_1$, the estimate of $t_0$ is increasing, while that of $z$ is decreasing.
The corrections are compatible with $t_1^{-1}$ and $t_1^{-2}$, respectively.
Fitting the results for $t_1 \ge 60$, not taking into account the statistical 
correlations, we arrive at $z=2.0244(4)$ and $t_0=-2.13(10)$.
As our preliminary estimate of this section we take
\begin{equation}
 t_0 =-2.1(2)   \;\;  ,  \;\;   z = 2.0245(10)   \;\;,
\end{equation}
which is compatible with both our extrapolation in $t_1$ and 
the result obtained for $t_1=160$. 

\begin{table}
\caption{\sl \label{fitz} 
Simulations with the Metropolis algorithm for $L=300$.
We give the results of minimizing the variance 
of $z_{m,eff,t_0}$ within the intervals $t_1  \le t < 2 t_1$, 
eq.~(\ref{mini}), with respect to $t_0$.
}
\begin{center}
\begin{tabular}{rll}
\hline
 \mc{1}{c}{$t_1$} &  \mc{1}{c}{$t_0$}  & \mc{1}{c}{$z$}  \\
\hline
20 &  -1.380(5) &   2.04089(20)  \\
30 &  -1.571(9)  &  2.03315(24)  \\
40 &  -1.679(13) &  2.03006(28) \\
60 &  -1.838(25) &  2.02677(36) \\    
80  & -1.944(38) &  2.02524(42) \\
120 & -1.95(8)  &   2.02523(54) \\
160 & -2.01(12) &   2.02481(65) \\ 
240 & -2.01(24) &   2.02470(87) \\
\hline
\end{tabular}
\end{center}
\end{table}

\begin{figure}
\begin{center}
\includegraphics[width=14.5cm]{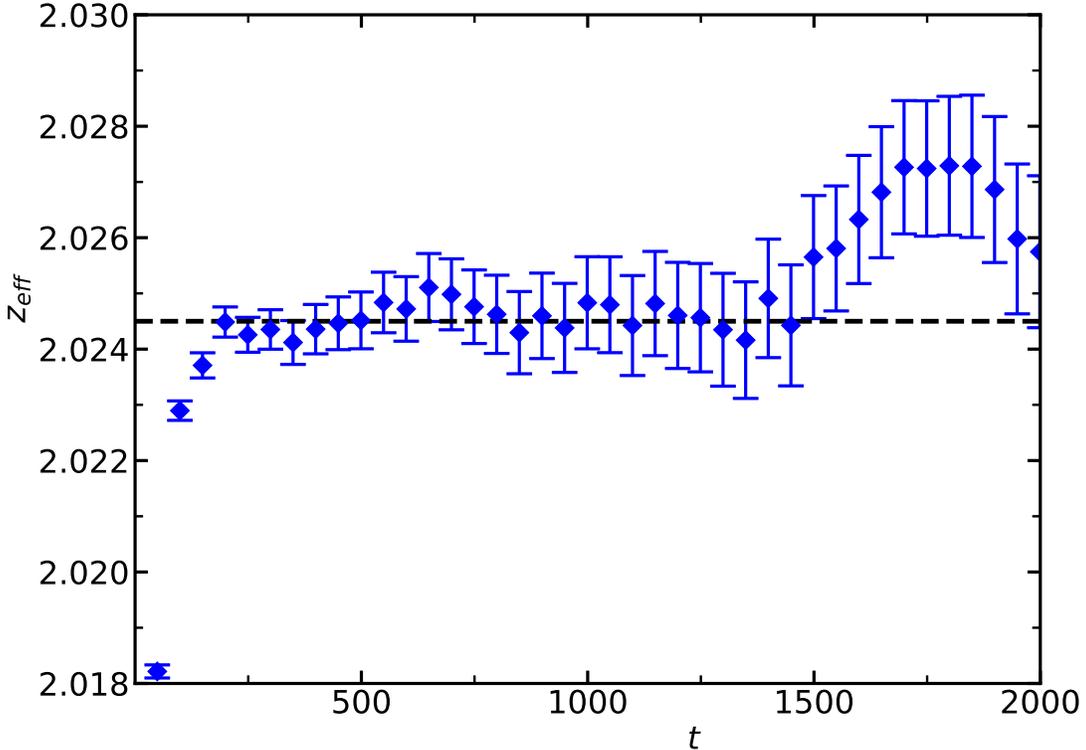}
\caption{\label{zeff300}
Simulations with the Metropolis algorithm.
The effective exponent $z_{eff}$ as defined by eq.~(\ref{zeff}) for $L=300$ and $t_0=-2.1$.
The dashed line indicates the preliminary result $z=2.0245$ of this section.
}
\end{center}
\end{figure}

As a check, in figure \ref{zeff300}, we plot $z_{m,eff,t_0}(t)$ for 
$t_0 = 3.1$ for the full range of $t$ that we have simulated.

\subsection{Simulations using the heat bath algorithm algorithm}
\label{HeatNER}
In this section we discuss simulations similar to those of the previous
one, replacing the Metropolis by the heat bath algorithm.  Details of 
the simulation program are discussed in the appendix \ref{SSE2program}.
Based on the results obtained above, we simulated lattices of the linear size
$L=300$. We run the simulations up to $t=2000$. 
We  performed 10000 runs with 32 replicas each. In total,
these simulations  took the equivalent of about 370 days on
a single core of a Intel(R) Xeon(R) CPU E3-1225 v3 CPU.

\begin{figure}
\begin{center}
\includegraphics[width=14.5cm]{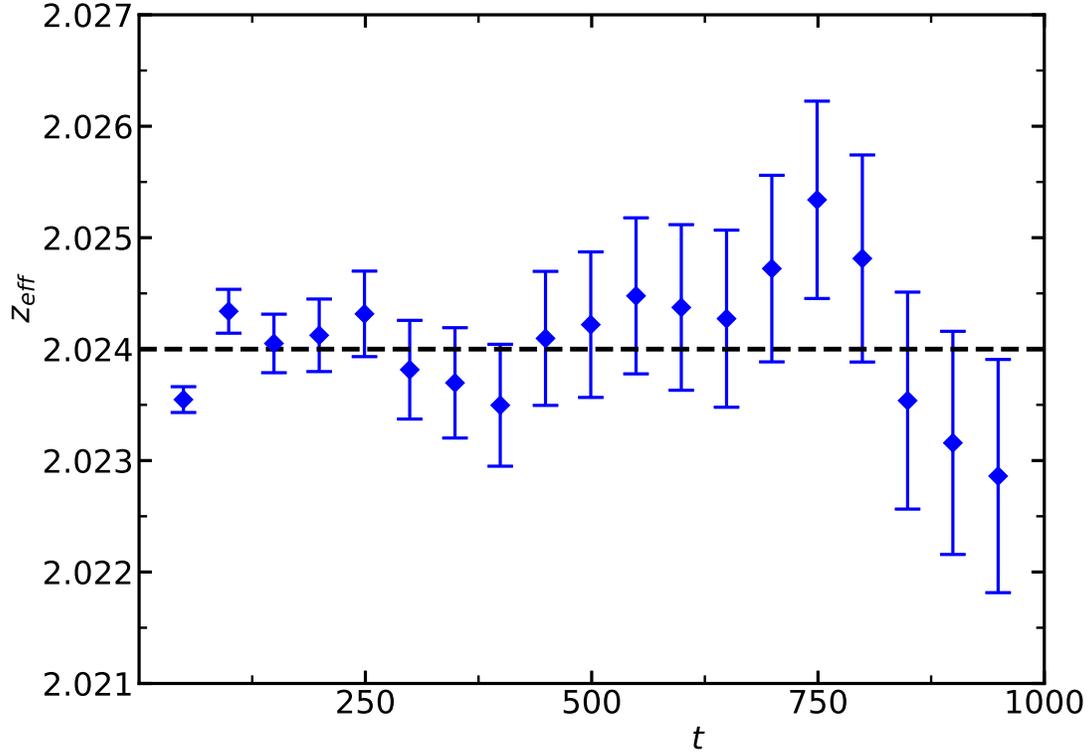}
\caption{\label{zeff300HB}
Simulations with the heat bath algorithm.
The effective exponent $z_{eff}$ as defined by 
eq.~(\ref{zeff}) for $L=300$ and $t_0=0$.
The dashed line indicates the preliminary result $z=2.024$ of this section.
}
\end{center}
\end{figure}

First we compared the relaxation times of the heat bath and the Metropolis
algorithm.  
We computed the ratio $t_M(m)/t_{HB}(m)$, where $t_M(m)$ and $t_{HB}(m)$ are the 
times needed by the Metropolis and the heat bath algorithm to reach a certain
value $m$ of the magnetization. Using a linear extrapolation in $t$ we arrive
at the estimate $t_M/t_B=1.3304(8)$ for the limit $t_B,t_M \rightarrow \infty$.
This ratio is in good agreement with the ratio of autocorrelation times, 
reported in table \ref{compare} above.

\begin{table}
\caption{\sl \label{fitzB} Same as table \ref{fitz} but for the 
heat bath instead of the Metropolis algorithm.
}
\begin{center}
\begin{tabular}{rll}
\hline
 \mc{1}{c}{$t_1$} &  \mc{1}{c}{$t_0$}  & \mc{1}{c}{$z$}  \\
\hline
15  & \phantom{-}0.0148(25) &   2.02379(14) \\
20  & \phantom{-}0.0219(49) &   2.02421(18) \\
30  & \phantom{-}0.031(8)   &   2.02459(24) \\
40  & \phantom{-}0.010(13)  &   2.02395(29) \\
60  & -0.003(27)            &   2.02364(40) \\
80  & \phantom{-}0.026(44)  &   2.02409(49) \\
120 & \phantom{-}0.032(89)  &   2.02420(66) \\
160 & \phantom{-}0.09(15)   &   2.02467(82)  \\
\hline
\end{tabular}
\end{center}
\end{table}

In table \ref{fitzB} we report our result for $z$ and $t_0$ obtained 
from the minimization procedure discussed above for the Metropolis algorithm.
Compared with the estimates
reported in table \ref{fitz} for the Metropolis algorithm, the estimates
for $z$ and $t_0$  show very little dependence on the range in $t$.
Therefore we abstain from extrapolating the results. Based on the result
for $t_1=40$ we take $z=2.0240(8)$ and $t_0=0.0(1)$ as our preliminary
result. The error bars are taken such that they include all estimates
with their error bars up to $t_1=120$.  As a check, in figure 
\ref{zeff300HB} we give $z_{eff}$, eq.~(\ref{zeff}), for $t_0=0$.

\section{Summary and Conclusions}
We have studied a purely dissipative relaxational dynamics for the 
improved Blume-Capel model 
on the simple cubic lattice. This model shares the universality  class of the 
three-dimensional Ising model. Improved means that the parameter $D$ of the 
model is chosen such that the amplitude of leading corrections to scaling 
is strongly suppressed. 
In particular, since we have to face critical slowing down when studying 
a relaxational process, it is important to use an improved model, since here 
already from relatively small lattices reliable results can be obtained. 

The numerical results for the dynamic critical
exponent $z$ given in the literature vary considerably.  In particular there 
is a clear discrepancy between most of the results obtained by the simulation
of the Ising model and field theoretic results. Only a previous simulation 
of the Blume-Capel model gives a result that is consistent with field theory.
 
We have computed the dynamic critical exponent by using two different 
approaches. As our final estimate we quote $z=2.0245(15)$
obtained from the finite size scaling analysis of equilibrium autocorrelation 
times at the critical point. The results that we obtain from the 
sudden quench of a fully magnetized configuration to criticality are fully 
consistent with this estimate.

Note that our estimate of the dynamic critical exponent is in nice agreement with
recent results obtained with the functional renormalization group method
\cite{Me15,FRG}. The same holds for the analysis of the four-loop
$\epsilon$-expansion \cite{Ad17} presented here in appendix \ref{analysing}.

\section{Acknowledgement}
I like to thank M. V. Kompaniets for very helpful correspondence. 
This work was supported by the Deutsche Forschungsgemeinschaft (DFG) 
under grant No HA 3150/5-1.

\appendix

\section{Parallel program using SSE2 intrinsics and shared use of 
random numbers}
\label{SSE2program}
In a second stage of the project we made an effort to speed up  our
simulation program for the heat bath algorithm. To this end we have chosen
an approach that is less involved than the multispin coding technique that
builds on bitwise operations. Here we exploited the \verb+SSE2+ instruction
set of x86 CPUs. These were accessed by using \verb+SSE2+ intrinsics.
\verb+SSE2+ instructions act on several variables that are packed into
128 bit units in parallel. In our case we store a single spin as
a 8 bit \verb+char+ variable, of which 16 are packed into a 128 bit unit.
To this end, we run 16 replicas of the system in parallel.
For each site $x$ we pack the 16 spins $s_x^{(j)}$,
where the upper index labels the replica, into a \verb+__m128i+ variable.
Computing the sum of the neighbors for the update and the measurement
of the magnetization are done in parallel for the 16 replicas. The actual
heat bath update is still done one by one.

Since the generation of a pseudo random number is relatively expensive,
it is a natural question, whether we can use the same stream of random numbers
for several replicas. One simple idea is to take a stream $r_i^{(0)}$
of random numbers that are uniformly distributed in $[0,1)$ and then use
the family 
\begin{equation}
\label{familyR}
r_i^{(j)} = \mbox{frac}(r_i^{(0)} + j/N) \; , 
\end{equation}
where $j=0,1,2,...,N-1$ for the simulation of $N$ replicas and $\mbox{frac}$
is the fractional part of a real number.
This way, all replicas are simulated by using a well behaved pseudo random
number. However a statistical correlation among the replicas arised. 
We computed statistical errors by using the jackknife method. 
Not to corrupt the estimate of the statistical error, correlated replicas are 
always put in the same bin. For the simplicity of the program, we did not
measure the correlation of the runs that share the family of random numbers.
Instead, for a few lattice sizes we performed runs, where each replica has its
own pseudo random number. Then we compared the statistical errors obtained 
for equal statistics.
It turns out that in the case of the simulations in equilibrium at $\beta_c$
we see virtually no effect on the statistical error up to about $N=4$. 
For larger
values of $N$ we see a gradual increase of the relative statistical error with 
increasing $N$.  In our simulations reported in section \ref{main}
we use $N=4$ throughout.

In the case of the sudden quench from zero temperature to the critical one
we also experimented with eq.~(\ref{familyR}). 
For small $N$ we even find a small reduction
of the statistical error compared with independent random numbers. 

However it turned out that an even larger variance reduction can be obtained 
by sharing the random number in a different way.
P. Grassberger \cite{private} pointed out
that the heat bath algorithm applied to the Blume-Capel model fulfills
the property of "monotonicity" \cite{Con}; See also the introduction
of ref. \cite{Grassberger}.

Two replica $A$ and $B$ of the system are simulated. At time $t=0$
\begin{equation}
\label{unequal}
 s_x^{(A)} \ge s_x^{(B)}
\end{equation}
holds for all sites $x$, where the upper index denotes the replica.
These two replicas are simulated by running through the sites in
the same order, using exactly the same stream of random numbers for both
systems. Then the condition~(\ref{unequal}) is preserved by the update.
This property is actually easy to prove. Let us start with the precise
definition of the heat bath update:
At the site $x$ the new value of the spin $s_x$
is chosen with the probabilities
\begin{equation}
p(-1) = \exp(-D-\beta S_x)/z \;\;, \; p(0) = 1/z \;\;, \;
p(0) = \exp(-D + \beta S_x)/z
\end{equation}
with $z=\exp(-D-\beta S_x)+1+\exp(-D+\beta S_x)$ and
$S_x =\sum_{y.nn.x} s_y$ is the sum of the nearest neighbors.
This is implemented in the following way:
First a uniformly distributed random number $r \in [0,1)$ is drawn.
Then $s_x' =-1$ is taken if $r<p(-1)$, $s_x' =0$, if $p(-1) \le r<p(-1)+p(0)$
and  $s_x' =1$, if $p(-1)+p(0)  \le r$. Since $p(-1)$ is monotonically
decreasing with increasing $S_x$, it follows for any given $r$ that
if $S_x^{(A)} \ge S_x^{(B)}$  then $s_x'^{(A)} \ge s_x'^{(B)}$.
Hence starting with eq.~(\ref{unequal}) this property is preserved throughout
the simulation.
This property allows for a variance reduction in the measurement of time 
dependent magnetization densities and variances thereof.
For details see ref. \cite{Grassberger}.

In our case, we run two replicas initialized with positive and negative 
magnetization, using the same stream of random numbers. The estimator 
of the magnetization is
\begin{equation}
m_I(t)  = \frac{m_{+}(t)-m_{-}(t)}{2}  \;,
\end{equation}
where the subscript $\pm$ indicates the initialization of the system.
By construction $m_I(t) \ge 0$ for all $t$. Hence at least for large
$t$, when the system is close to equilibrium, there should be a reduction 
of the variance. The numerical experiment shows that this is also the 
case for times $t$ relevant in our study.
We compare with two systems running with independent random number
streams.
We find that initially the gain increases rapidly from about 1.3
for the first measurement to about 2.2  at $t\approx 30$. Then it slowly
increases up to about 2.5 at $t=2000$. 

We also tried to combine this idea with eq.~(\ref{familyR}). Unfortunately
we see a counteracting effect. Taking also into account the CPU time needed
to generate a random number, we have chosen $N=2$ for our production runs.
In our \verb+SSE2+ program we have simulated in parallel 16 replica with all 
spin up and 16 replica with all spin down initialization using 8 independent
streams of pseudo random numbers.

The general ideas on the reuse of random numbers and variance reduction 
are widely used. One can easily convince oneself by typing the keywords
"common random numbers" or "antithetic variates" in a search engine or 
have a look at a text book on Monte Carlo methods such as ref. \cite{Kroese}
for example.

\section{The two-dimensional Ising model}
\label{appendixA}  In the analysis of the four-loop 
$\epsilon$-expansion result \cite{Ad17}
we shall use the numerical estimate of $z$ for the universality class 
of the two-dimensional Ising model as boundary condition. The most precise
estimates given in the literature are $z=2.1665(12)$ \cite{NiBl96} and 
$z= 2.1667(5)$  \cite{NiBl00}.  These estimates  were obtained from 
the analysis of very accurate estimates of $\tau_{exp}$
 obtained for linear lattices sizes $L \le 15$.  The
accuracy of the estimates of $z$ relies on the correctness of the ansatz for 
corrections to scaling. The leading correction is proportional to $L^{-2}$. 
In refs. \cite{NiBl96} and  \cite{NiBl00} also subleading corrections had
to be taken into account.

Here we performed simulations of the two-dimensional Ising model on the 
square lattice exactly at the critical temperature. 
We determine the integrated autocorrelation time of the magnetic
susceptibility in exactly the same way as in section \ref{NumTC}.  
We performed simulations for 27 different linear lattice sizes from 
$L=8$ up to $120$.

We fitted our data using the ans\"atze 
\begin{eqnarray}
\label{z0}
\tau  & = & c L^{z} \;, \\
\label{z1}
\tau  & = & a L^{z} \; (1 + b L^{-2}) \;, \\
\label{z2}
\tau  & = & a L^{z} \; (1 + b L^{-2} + c L^{-4}) \;.
\end{eqnarray}
For example using the ansatz~(\ref{z2}) we get 
$\chi^2/$d.o.f.$=1.13$, $a=0.06417(4)$, $b=6.42(5)$, $c=-17.4(2.9)$ and
$z=2.1663(2)$, when including all lattice sizes $L\ge10$ in the analysis.
Using the ansatz~(\ref{z1}) we get 
$\chi^2/$d.o.f.$=1.00$, $a=0.06397(9)$, $b=6.89(20)$ and $z=2.1670(4)$
taking into account all lattice sizes $L\ge16$. Using the ansatz without 
corrections we get $\chi^2/$d.o.f.$=0.82$, $a=0.0644(2)$ and $z=2.1657(7)$ 
using all data with $L\ge56$.  Assessing all fits that we performed, 
we arrive at the estimate $z=2.167(2)$, confirming the results of  
refs. \cite{NiBl96,NiBl00}. 

\section{Analyzing the field theoretic results}
\label{analysing}
Here we make an attempt to extract a number for $z$ for three 
dimensions by using the four-loop $\epsilon$-expansion result \cite{Ad17}.  
For the Ising universality class the authors give 
\begin{equation}
\label{fourloop}
z= 2 + 0.0134461561 \epsilon^2+ 0.011036273(10) \epsilon^3
- 0.0055791(5) \epsilon^4  +O(\epsilon^5) \; .
\end{equation}
Reexpressing this result by using eq.~(\ref{splitz})  we get
\begin{equation}
\label{cexpansion}
c = 0.72609243 \; \left (1 - 0.188484 \epsilon + 0.22506 \epsilon^2 + ...
\right) \;.
\end{equation}
The $[1/1]$ Pad\'e approximation is   
\begin{equation}
\label{pade}
c \approx 0.72609243 \;\; \frac{1+1.00557  \epsilon}{1+1.19405 \epsilon} \;.
\end{equation}
Inserting $\epsilon=1$ and $2$, using $\eta=0.0362978(20)$ 
\cite{Simmons-Duffin:2016wlq}
and $\eta=1/4$ we get $z=2.0241$ and $2.1613$, respectively.
Enforcing $z=2.167$ in two dimensions, we arrive at
\begin{equation}
\label{pade2}
c \approx 0.72609243 \;\; \frac{1+0.82727  \epsilon+0.03361 \epsilon^2}
  {1+ 1.01575 \epsilon} \;
\end{equation}
resulting in $z=2.0243$ for three dimensions.

Bausch et al. \cite{Bausch81} studied the dynamics of an interface
in $1+ \epsilon'$ dimensions. They arrive at 
$z = 2 + \epsilon' - \frac{1}{2} \epsilon'^2 + ...$
for the dynamic critical exponent. They give an interpolation of their
result and the two-loop $\epsilon$-expansion, eq.~(9) of \cite{Bausch81}. 
Inserting $d=2$ and $3$, one gets $z=2.126$ and $2.019$, respectively.

Extending the approach of  \cite{Bausch81} by using the result of 
 \cite{Ad17} we arrive at
\begin{equation}
\label{bauschres}
z-2 \approx \frac{(0.0348932 - 0.0076607 d) \; (d-1) \; (4-d)^2 }
           {1-1.532616 d + 0.910303 d^2 -  0.132595 d^3} \;.
\end{equation}
Inserting $d=2$ and $3$, one gets $z=2.1519$ and $2.0235$, respectively.
Enforcing $z=2.167$ in two dimensions, we arrive at
\begin{equation}
\label{bauschres2}
z-2 \approx \frac{(0.0419521 - 0.0131387 d + 0.00083183 d^2) \; (d-1) \; (4-d)^2 }
           {1 - 1.443329 d + 0.834824 d^2 -0.124687 d^3}
\end{equation}
giving $z=2.0245$ in three dimensions.

The four-loop result for the expansion in three dimensions 
fixed is \cite{Prud97}:
\begin{equation}
z-2 = 0.008399 g^2 -0.000045 g^3 -0.020423 g^4  \;.
\end{equation}
Following the idea of ref. \cite{HaHoMa72} we might analyze
\begin{equation}
\frac{z-2}{\eta} = 0.765359 (1 - 0.088666 g + 2.275305 g^2) \;,
\end{equation}
where the series for $\eta$ is taken from ref. \cite{gstar}, eq.~(2.4).
We arriving at the $[1/1]$ Pad\'e approximation
\begin{equation}
\frac{z-2}{\eta} \approx  0.765359 \frac{1 + 25.5729 g}{1 + 25.6615 g} \;.
\end{equation}
Inserting the fixed point value $g^*=1.4299$, eq.~(22) of ref. \cite{Prud97},  and 
$\eta=0.0362978(20)$, we get $z=2.0277$, which is considerably larger than
the value obtained by the Pad\'e approximation for $z-2$ itself. 

We find that the estimates obtained by different  resummation 
schemes scatter less for the four-loop $\epsilon$-expansion than
for the four-loop expansion in three dimensions fixed. As our final 
estimate we take 
\begin{equation}
 z = 2.0243
\end{equation}
from eq.~(\ref{pade2}). Assigning an error bar is a difficult task. Comparing
the different estimates  
eqs.~(\ref{pade},\ref{pade2},\ref{bauschres},\ref{bauschres2}), it should
be at most a one in the third decimal place. 

\section{Leading corrections to scaling}
\label{leadingC}
\begin{figure}
\begin{center}
\includegraphics[width=14.5cm]{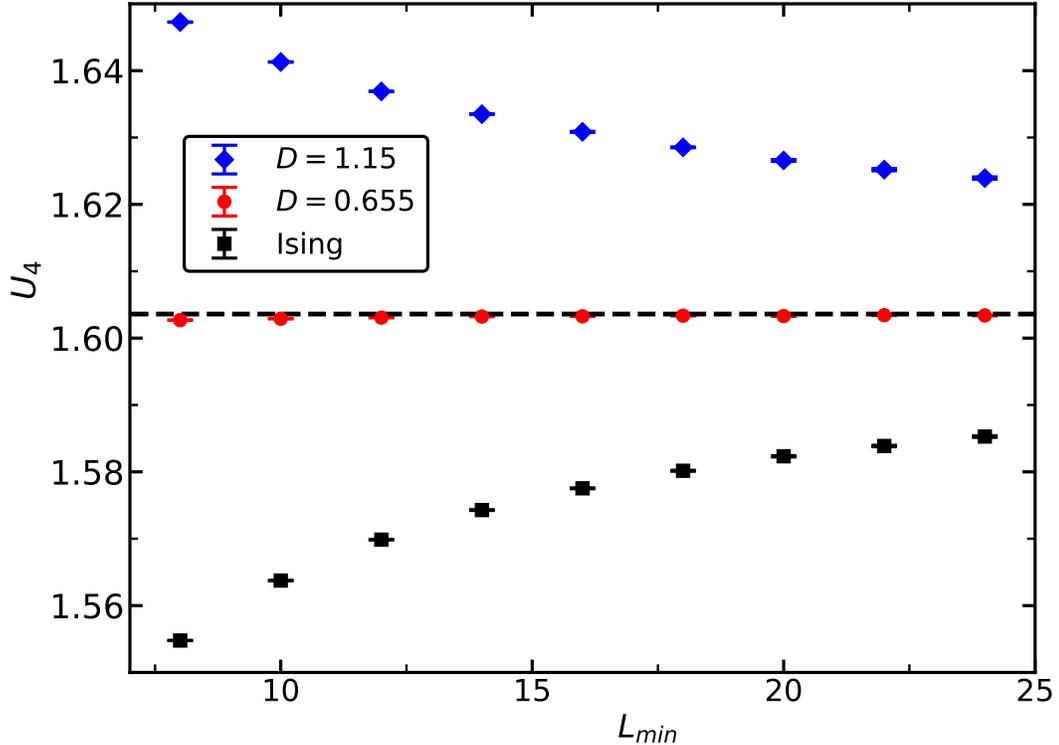}
\caption{\label{binderplot}
We plot  the Binder cumulant 
$U_4=\frac{\langle m^4 \rangle}{\langle m^2 \rangle^2}$ at the 
critical temperature for the Ising model and the Blume-Capel model at
$D=0.655$ and $1.15$. Note that the error bars are clearly smaller than the 
symbol size. The dashed line gives the estimate 
$U_4^* = 1.6036(1)$ of the fixed point value \cite{myBC}. 
}
\end{center}
\end{figure}

In order to study the effect of leading corrections to scaling on the 
autocorrelation times, we simulated the Ising model and the Blume-Capel model
at $D=1.15$ at the estimates of the inverse critical  temperature
$\beta_c=0.221654626(5)$, ref. \cite{pushing}, and $0.4756110(2)$, 
ref. \cite{myBC}, respectively.   We simulated 
the linear lattice sizes $L=8, 10, 12, ..., 24$
by using the heat bath algorithm with checkerboard decomposition.
In order to demonstrate the size of corrections to scaling, we plot in 
figure \ref{binderplot} the Binder cumulant $U_4=\frac{\langle m^4 \rangle}
{\langle m^2 \rangle^2}$ for the Ising model and the Blume-Capel model at
$D=0.655$ and $1.15$. At the critical point it behaves as
\begin{equation}
 U_4(L,D) = U_4^* + a(D) L^{-\omega} + b \; a^2(D) L^{-2 \omega} + ...
 + c(D) L^{-\omega'} + ... \;,
\end{equation}
where the term $c(D) L^{-\omega'}$ represents subleading corrections.
Almost degenerate subleading correction exponents are $2-\eta=1.9637022(20)$ due 
to the analytic background in the magnetic
susceptibility and $\omega_{NR}=2.022665(28)$ 
\cite{Simmons-Duffin:2016wlq}.
Estimates of the fixed point value are $U_4^* = 1.6036(1)$, ref. \cite{myBC}
and $1.60356(15)$,  ref. \cite{pushing}.

Fitting the data for $U_4$ at $D=0.655$ confirms that the amplitude of leading 
corrections vanishes at the level of our numerical precision. Next we analyzed
the ratios
\begin{equation}
r_{U_4}(D,L) = \frac{U_4(D,L)}{U_4(D=0.655,L)}
\end{equation}
using the ans\"atze
\begin{equation}
\label{RU1}
r_{U_4}(D,L) = 1 + a_U(D) L^{-\omega} 
\end{equation}
and
\begin{equation}
\label{RU2}
r_{U_4}(D,L) = 1 + a_U(D) L^{-\omega} + b_U(D) L^{-2} \;,
\end{equation}
where we have fixed  $\omega=0.82968$. 

For the Ising model we find
using the ansatz~(\ref{RU2}), including all data $a_U(-\infty)=-0.1530(5)$, 
$b_U(-\infty)=-0.178(8)$ and $\chi^2/$d.o.f.$=0.68$. Instead using 
the ansatz~(\ref{RU1}) we get $a_U(-\infty)=-0.1577(5)$ and 
$\chi^2/$d.o.f.$=0.8$ when including all lattice sizes with $L\ge 20$. 
We conclude $a_U(-\infty)=-0.155(4)$.

For $D=1.15$, using the ansatz~(\ref{RU2}), including all data with $L\ge 10$
we get $a_U(1.15)=0.1839(5)$, $b_U(1.15)=-0.328(9)$ and $\chi^2/$d.o.f.$=1.22$.
Instead, using the ansatz~(\ref{RU1}) we get 
$a_U(1.15)=0.1760(8)$ and $\chi^2/$d.o.f.$=2.83$ 
when including all lattice sizes with $L\ge 20$. We conclude 
$a_U(1.15)=0.180(5)$.

Next we analyzed ratios of susceptibilities
\begin{equation}
r_{\chi}(D,L) = \frac{\chi(D,L)}{\chi(D=0.655,L)} \;,
\end{equation}
where the powerlike divergence $\chi \propto L^{2-\eta}$ cancels.
We fitted these ratios with
\begin{equation}
\label{rfix1M}
r_{\chi}(D,L) = c_{\chi}(D) (1 + a_{\chi}(D) L^{-\omega} )
\end{equation}
and as check
\begin{equation}
\label{rfix2M}
r_{\chi}(D,L) = c_{\chi}(D) \; (1 + a_{\chi}(D) L^{-\omega}  + b_{\chi}(D) L^{-2})  \; ,
\end{equation}
where we have fixed  $\omega=0.82968$.
In the case of the Ising model we get $\chi^2/$d.o.f.$=0.84$ and
$a_{\chi}(-\infty)=-0.2211(7)$ including all data with $L \ge 10$ using 
the ansatz~(\ref{rfix1M}). Instead, using the ansatz~(\ref{rfix2M}) we 
get $\chi^2/$d.o.f.$=1.00$, $a_{\chi}(-\infty)=-0.220(6)$ and 
$b_{\chi} (-\infty)= -0.01(5)$, taking into account $L \ge 10$. 
We take $a_{\chi}(-\infty)=-0.220(6)$ as our final result.

In the case of 
$D=1.15$ we get by using the ansatz~(\ref{rfix1M}) taking into
account all data for $L\ge 12$ the result $\chi^2/$d.o.f.$=1.13$ and
$a_{\chi}(1.15)=0.2483(15)$.
Instead, using the ansatz~(\ref{rfix2M}) we
get $\chi^2/$d.o.f.$=0.70$, $a_{\chi}(1.15)=0.262(7)$ and 
$b_{\chi} (1.15)=-0.13(6)$, taking into account $L \ge 10$. As our final 
estimate we take $a_{\chi}(1.15)=0.255(14)$ that covers both fits, 
including their error bars.

Finally we computed ratios of autocorrelation times
\begin{equation}
r_{\tau}(D,L) = \frac{\tau(D,L)}{\tau(D=0.655,L)} \;,
\end{equation}
where the power divergence $\propto L^z$ should cancel and, 
hopefully also corrections due to the breaking of the Galilean symmetries
by the lattice to a large extend. We fitted these ratios by using 
the ans\"atze 
\begin{equation}
\label{rfix1}
r_{\tau}(D,L) = c_{\tau}(D) \; (1 + a_{\tau}(D) L^{-\omega} ) 
\end{equation}
and as check
\begin{equation}
\label{rfix2}
r_{\tau}(D,L) = c_{\tau}(D) \; (1 + a_{\tau}(D) L^{-\omega}  + b_{\tau}(D) L^{-2})  \; ,
\end{equation}
where we have fixed  $\omega=0.82968$.

In the case of the Ising model we get from~(\ref{rfix1}), including 
all data with $L\ge 10$ the estimate $a_{\tau}(-\infty) = -0.452(2)$
and $\chi^2/$d.o.f.$=1.00$. Fitting with the ansatz~(\ref{rfix2}), 
$b_{\tau}(D)$ is compatible with zero and $a_{\tau}(-\infty) = -0.43(2)$. 
We conclude  $a_{\tau}(-\infty) = -0.44(3)$. 
For $D=1.15$ we get, fitting all data with $L\ge 14$ by using the 
ansatz~(\ref{rfix1}) the estimate $a_{\tau}(1.15)=0.602(10)$ and 
$\chi^2/$d.o.f.$=1.14$.  Fitting with the ansatz~(\ref{rfix2}),
using all data we get $a_{\tau}(1.15)=0.631(13)$, $b_{\tau}(1.15)=-0.49(8)$
and $\chi^2/$d.o.f.$=1.86$.  We conclude  $a_{\tau}(1.15)=0.62(3)$.
These fits support the hypothesis that $z$ does not depend on $D$, and 
the differences can be explained by corrections.

According to the renormalization group, leading corrections to scaling 
are caused by a unique scaling field. Therefore, the ratios of correction
amplitudes for different quantities assume universal values. In particular, 
for the improved model the amplitude of leading corrections vanishes for all
quantities.

For the susceptibility and the Binder cumulant we get
$a_{\chi}(-\infty)/a_{U}(-\infty) = $ \\ $[-0.220(6)]/[-0.155(4)]=1.42(7)$
and
$a_{\chi}(1.15)/a_{U}(1.15) = [0.255(14)]/[0.180(5)]= 1.42(9)$. We conclude
$a_{\chi}/a_{U}=1.42(9)$. 

For the autocorrelation time and the Binder cumulant we get 
\begin{equation}
 \frac{a_{\tau}(-\infty)}{a_{U}(-\infty)} = \frac{-0.44(3)}{-0.155(4)} 
= 2.84(30)
\end{equation}
and  
\begin{equation}
 \frac{a_{\tau}(1.15)}{a_{U}(1.15)} = \frac{0.61(3)}{0.180(5)} = 3.39(25)  \;,
\end{equation}
confirming the universality of the ratio of correction amplitudes. As 
our final result we take $a_{\tau}/a_{U} = 3.1(6)$.

Note that the amplitude of the leading correction is relatively large
for the autocorrelation time compared with the Binder cumulant and the 
magnetic susceptibility.  This might explain the wide spread of the estimates
of $z$ obtained from simulations of the three-dimensional Ising model, 
when the leading 
correction to scaling is not explicitly taken into account in the analysis.

\def\refname{}

\end{document}